\documentclass[twocolumn,showpacs,preprintnumbers,amsmath,amssymb,APS]{revtex4}

\usepackage{graphicx}
\usepackage{dcolumn}
\usepackage{bm}

\newcommand{\De}{\Delta} 
 
\newcommand{\be}{\beta} 
\newcommand{\beD}{\beta_\text{D}} 
\newcommand{\al}{\alpha} 
\newcommand{\ga}{\gamma}

\newcommand{\beq}{\begin{eqnarray}} 
\newcommand{\enq}{\end{eqnarray}}

\newcommand{\eps}{\varepsilon}

\newcommand{\Ham}{\begin{cal} H \end{cal}}
\newcommand{\Ord}{\begin{cal} O \end{cal}}
\newcommand{\si}{\sigma} 
\newcommand{\thi}{\theta} 
\newcommand{\sT}{\sigma^\text{(T)}}
\newcommand{\sR}{\sigma^\text{(R)}}
\newcommand{\FT}{F^\text{(T)}}
\newcommand{\dsim}{\delta_{\sigma_i,-1}}
\newcommand{\dsip}{\delta_{\sigma_i,1}}
\newcommand{\dthim}{\delta_{\theta_i,-1}}
\newcommand{\dthip}{\delta_{\theta_i,1}}
\newcommand{\dRm}{\delta_{\sigma_i^\text{(R)},-1}}
\newcommand{\dRp}{\delta_{\sigma_i^\text{(R)},1}}
\newcommand{\dTm}{\delta_{\sigma_i^\text{(T)},-1}}
\newcommand{\dTp}{\delta_{\sigma_i^\text{(T)},1}}

\newcommand{\gaPP}{\gamma_\text{PP}}
\newcommand{\gaHH}{\gamma_\text{HH}}
\newcommand{\gaHP}{\gamma_\text{HP}}
\newcommand{\bD}{\beta_\text{D}} 
\newcommand{\PD}{P_\text{D}} 
\newcommand{\ZD}{Z_\text{D}}

\newcommand{\NHT}{N_\text{H}^\text{(T)}} 
\newcommand{\NHR}{N_\text{H}^\text{(R)}} 
\newcommand{\nHT}{n_\text{H}^\text{(T)}}

\newcommand{\om}{\omega}

\newcommand{\Om}{\Omega}

\newcommand{\muNHT}{\mu_{N_\text{H}^\text{(T)}}}
\newcommand{\xiNHT}{\xi_{N_\text{H}^\text{(T)}}}

\newcommand{\fopt}{f_\text{opt}}

\newcommand{\fmax}{f_\text{max}}

\newcommand{\sal}{\sigma^{(\alpha)}}
\newcommand{\Fal}{F^{(\alpha)}}

\newcommand{\fsTth}{f_{\sigma^\text{(T)},\theta}}

\newcommand{\fsT}{f_{\sigma^\text{(T)}}}
\newcommand{\lsTth}{l_{\sigma^\text{(T)},\theta}}
\newcommand{\lsT}{l_{\sigma^\text{(T)}}}

\newcommand{\WPPsTth}{{W}^\text{PP}_{\sigma^\text{(T)},\thi}}
\newcommand{\WHHsTth}{{W}^\text{HH}_{\sigma^\text{(T)},\thi}}
\newcommand{\WHPsTth}{{W}^\text{HP}_{\sigma^\text{(T)},\thi}}

\newcommand{\WPPnHT}{{W}^\text{PP}_{n_H^\text{(T)}}}
\newcommand{\WHHnHT}{{W}^\text{HH}_{n_H^\text{(T)}}}
\newcommand{\WHPnHT}{{W}^\text{HP}_{n_H^\text{(T)}}}
\newcommand{\DPPoWnHT}{{D}^\text{PP}_{n_H^\text{(T)}}}
\newcommand{\DHHoWnHT}{{D}^\text{HH}_{n_H^\text{(T)}}}
\newcommand{\DHPoWnHT}{{D}^\text{HP}_{n_H^\text{(T)}}}
\newcommand{\DPPoWsTth}{{D}^\text{PP}_{\sigma^\text{(T)},\thi}}
\newcommand{\DHHoWsTth}{{D}^\text{HH}_{\sigma^\text{(T)},\thi}}
\newcommand{\DHPoWsTth}{{D}^\text{HP}_{\sigma^\text{(T)},\thi}}
\newcommand{\WPP}{{W}^\text{PP}}
\newcommand{\WHH}{{W}^\text{HH}}
\newcommand{\WHP}{{W}^\text{HP}}
\newcommand{\DPP}{{D}^\text{PP}}
\newcommand{\DHH}{{D}^\text{HH}}
\newcommand{\DHP}{{D}^\text{HP}}

\begin{document}

\title{Dry and wet interfaces: Influence of solvent particles on molecular recognition}  

\author{Johannes Taktikos}
\author{Hans Behringer}
\affiliation{Fakult\"at f\"ur Physik, Universit\"at Bielefeld, 33615 Bielefeld, Germany}%

\begin{abstract}
  We present a coarse-grained lattice model to study the influence of
  water on the recognition process of two rigid proteins. 
  The basic model is formulated in terms of the hydrophobic effect. We
  then investigate several modifications of our basic model showing
  that the selectivity of the recognition process can be enhanced by
  considering the explicit influence of single solvent particles. When
  the number of cavities at the interface of a protein-protein complex
  is fixed as an intrinsic geometric constraint, there typically
  exists a characteristic fraction that should be filled with water
  molecules such that the selectivity exhibits a maximum. In addition
  the optimum fraction depends on the hydrophobicity of the interface
  so that one has to distinguish between dry and wet interfaces.
\end{abstract}

\pacs{87.15.A, 87.15.-v ,89.20.-a}

\maketitle

\section{\label{sec:Intro} Introduction \protect}

Molecular recognition denotes the ability of a certain biomolecule to
find the right partner molecule in an heterogeneous environment, such
that the formed complex can perform its assigned biological task.
Prominent examples of specific recognition processes between proteins
comprise enzyme-substrate binding, antigen-antibody binding or
protein-receptor interactions \cite{Alberts1994, Kleanthous2000}.  It
is a remarkable property of recognition processes that a biomolecule
(called probe molecule throughout this article) can identify its
"correct" complex partner by distinguishing between the supposed
"target" and a competing "rival" molecule that possibly features only
a slightly different structure at the binding epitope.  Therefore, an
understanding of molecular recognition processes is obviously not only
interesting from a biological point of view, but also necessary for
various biotechnological or pharmaceutical applications.  The high
specificity of molecular recognition processes can be illustrated by
the "lock-and-key" mechanism for inflexible biomolecules which demands
a high geometrical complementarity for the two molecules forming a
complex \cite{Fischer1894,Pauling1940}. For that reason, there is in
general only one possible binding partner (say "key") for a given
molecule ("lock"). As most macromolecules prove to be flexible, the
so-called "induced-fit" scheme has been established, according to
which the necessary complementarity is only achieved after some
conformational changes of the corresponding backbones of the proteins
\cite{Koshland1958}.

The forces that stabilize a protein complex basically emerge from a
complicated interplay between non-covalent bonds. These bonds are
characterized by energies of the order of $2-6$ kcal/mol
\cite{Flyvbjerg1997}. Since this is only slightly stronger than the
thermal energy $k_\text{B} T_\text{room} \approx 0.62$ kcal/mol at
physiological conditions, we can conclude that the formation of a
stable protein complex demands a large number of non-covalent bonds
and thus many participating functional groups with appropriate
complementarity \cite{Pauling1940}. It has been investigated that the
driving forces for molecular recognition are dominated by hydrogen
bonds and especially by the hydrophobic effect
\cite{Jones1996,Larsen1998,Wodak2003,Kleanthous2000,Janin2007}.  The
hydrophobic effect sums up the mechanism that the hydrophobic residues
of proteins are effectively pushed together when the polar solvent
leaves the space between the hydrophobic amino acids for entropical
and energetical reasons \cite{Jackson2006}.

The enormous significance of water for biological systems has been
manifest for many years \cite{Levy2006}. Although water is essential
for the structure, stability, dynamics and functions of biomolecules,
biological models often describe the solvent only as a passive
component of the system as is done, for example, by referring to the
hydrophobic effect. However, it has been shown, that water molecules
which are imbedded in cavities between two bounded proteins play a
crucial role for the formation and stabilization of the complex and
can thus be considered as an active part of the structure
\cite{Covell1997,Cheung2002,Papoian2003,Wodak2003,Rodier2005,Levy2006,Li2007}.
Indeed it has been observed that in interfaces between two proteins
about 10-20 \% of the area is made up of cavities on average of which
a large number are filled by at least one water molecule
\cite{Hubbard1994,Larsen1998,Sonavane2008}. The energetic
contributions of the buried water molecules are basically twofold.
They can either contribute van der Waals interactions with adjacent
amino acids or form hydrogen bonds between constituents of the two
proteins (sometimes involving more than one buried water molecule).
The latter possibility requires a high degree of geometric
directionality of the involved molecules and parts of the proteins.
The energetic contributions due to these mediated interactions are
typically smaller by a factor of two or three than direct contacts,
however, examples where they are of the same strength as direct
contacts do exist \cite{Covell1997,Papoian2003,Li2007}.

Interfaces of protein complexes show different levels of hydration and
can exhibit up to as many interactions caused by imbedded water
molecules as by direct hydrogen or salt bridges \cite{Rodier2005}. On
experimental grounds one can basically distinguish between "wet"
interfaces with many imbedded water molecules and "dry" interfaces
where water is absent \cite{Janin1999, Wodak2003,Rodier2005}. Dry
interfaces typically feature a ring of water molecules around the
binding epitope. In general the less hydrophobic interfaces between
antibodies and antigens tend to be wet whereas the more hydrophobic
protease-inhibitor interfaces appear to be dry. This suggests a
correlation between the hydrophobicity of the interface and the degree
of hydration. Note however that exceptions to this broad rule do
exist.

In this article we will investigate the influence of buried water
molecules in protein-protein interfaces on the selectivity of the
corresponding recognition process. Our considerations are carried out
within a coarse-grained approach where the bulk solvent degrees of
freedom are integrated out. The energetics is then formulated on the
level of amino acids in terms of the hydrophobic effect between
residues of different hydrophobicity. Additional residual water
degrees of freedom which are imbedded in the interface and can thus
actively mediate interactions between amino acids are then
incorporated into the model. From the point of view of modeling this
can be done by applying direct and water-mediated contact energies
\cite{Papoian2003} or by using generic double well potentials of mean
forces with one minimum corresponding to direct contacts of two
residues and a characteristic second one resulting from
water-separated contacts \cite{Cheung2002,Levy2006}. We finally remark
that the problem of molecular recognition has been considered in
coarse-grained approaches in several articles
\cite{Lancet1993,Janin1997,Rosenwald2002,Wang2003,Bogner2004,Behringer2006,Behringer2007a,Lukatsky2008,Behringer2007,Behringer2008}.

For our investigations we utilize a general two-stage approach (Sec.
\ref{sec:Modelling}) where in a first step an ensemble of probe
molecules is designed with respect to a given target. In a second
step, we investigate the recognition ability or selectivity of the
probe ensemble by comparing the associated free energy for the two
cases that the probe molecules bind the target or a different rival
molecule, respectively. In the subsequent sections, we will modify the
"elementary" hydrophobic-polar (HP) model by taking the direct
influence of single solvent molecules into account.  Nevertheless, we
have to keep in mind that the protein interaction with water is
already part of the HP model since its energetics are based on the
hydrophobic effect.  In the following sections we analyses the
influence of buried water molecules in the interface on the
selectivity of molecular recognition.  Whereas in Sec.
\ref{sec:BadContacts} every cavity at the interface is filled by a
water molecule, in Sec. \ref{sec:Option} we make the inclusion
optional and additionally couple the water's interaction to the
adjacent type of amino acid.  In particular, we will investigate
whether or not the inclusion of solvent molecules in the interface can
lead to an enhancement of the selectivity. The technical details how
the selectivity for the model with an optional inclusion of water is
calculated are discussed in the appendix.

\section{\label{sec:Modelling} General approach to molecular recognition \protect}

In this section we briefly discuss how we model the recognition
process and introduce a measure of its selectivity (more detailed
accounts can be found elsewhere \cite{Behringer2006, Behringer2007,
  Behringer2008}). We model a protein's recognition site at the
interface of a protein-protein complex as a two-dimensional array of
$N$ amino acids, also called residues or monomers. Typical values of
$N$ range between 30 and 60 \cite{Kleanthous2000}. For the description
of a so called probe molecule $\thi$, that is supposed to recognize a
certain target molecule, we introduce the $N$-dimensional vector $\thi
= (\thi_1,\ldots,\thi_N)$, whose $i$-th component indicates the type
of amino acid on site $i$. Accordingly, the target molecule $\si$ is
specified by its residues $\si = (\si_1,\ldots,\si_N)$. For the sake
of simplicity we assume that both proteins have the same number of
monomers at the interface which match when forming a complex. We note,
however, that systems where this assumption holds true do exist
\cite{Chapagain08}.

To specify a single residue one should a priori distinguish between
the 20 different amino acids occurring in nature. In the coarse-grained
approach of the hydrophobic-polar (HP) model, we reduce the alphabet
of amino acids to only two letters and differentiate between the polar
and non-polar (hydrophobic) subgroup. Thus we get an Ising-like
variable and choose the convention to attribute to $\si_i$ or
correspondingly $\thi_i$ the value $+1$ for a hydrophobic (H) and $-1$
for a polar (P) monomer at site $i$. We justify this procedure by
having in mind that hydrophobicity acts as the dominant driving force
in molecular recognition \cite{Kleanthous2000,Wodak2003,Janin2007}.
Furthermore one gets the two amino acid subgroups as a very good
approximation by applying an eigenvalue decomposition of the
Miyazawa-Jernigan matrix which consists of the pairwise interactions
between all natural amino acids \cite{Miyazawa1985, Li1997}. Note that
there exist also other methods to reduce the alphabet of amino acids
to five clustered subgroups \cite{Wang1999, Cieplak2001}.

The induced-fit theory motivates us to account for minor
rearrangements of amino acid side chains which provide the needed
complementarity for the formation of a protein-protein complex. This
feature is incorporated into the model by defining the quality of
contact between the binding partners, labeled as $S=(S_1,\ldots,S_N)$.
We just discriminate between "good" ($S_i = +1$) and "bad" ($S_i =
-1$) contacts at site $i=1,\ldots, N$. The (geometric) quality of the
contact can be understood as a characteristic trait of one of the
molecules or, alternatively, as a collective variable of the probe and
target molecule. The contact variable sums up all geometric conditions
at the interface, for example, the distances between opposite residues
or the alignment of their polar moments.  Its relevance for the
inclusion of water molecules at the interface is discussed in
paragraphs \ref{sec:BadContacts} and \ref{sec:Option}.

In our picture of the protein complex, we consider a general
Hamiltonian $\Ham(\si,\thi;S)$ depending on the structures $\si$ and
$\thi$ and some kind of interaction between binding partners at
position $i$ which is
related to the corresponding variable $S_i$.
We formulate the energetics at the interface by 
 a modified HP model \cite{Behringer2006}:
\begin{eqnarray}
\Ham(\si,\thi;S) := - \eps \sum_{i=1}^N \frac{1+S_i}{2}  \si_i  \thi_i. \label{HP}
\end{eqnarray}
The parameter $\eps > 0$ gives the strength of the hydrophobic
interaction and is typically of the order of 2 kcal/mol \cite{Chapagain08}. Note that the
factor $\frac{1+S_i}{2} \in \{0,1\}$ suppresses the contribution of
binding energy in the case of bad contacts. For a good contact at site
$i$ we receive the contribution $- \eps \si_i \thi_i $: If the type of
residues of the protein interface in contact is identical, i.e. $\si_i
\thi_i = 1$, we will get a favorable term $- \eps < 0$, whereas for
different types of amino acids the resulting $+ \eps$ represents a
non-favorable energy contribution.  We note that HP-like models have
been applied in various biophysical contexts over the last years
\cite{Go1983,Dill1985,Lau1989,
  Dill1995,Golumbfskie1999,Chakraborty2001,Polotsky2004a,Polotsky2004b,Bachmann2006}.

To study the recognition process between the two biomolecules, we
adopt a two-stage approach. In the first step, also referred to as the
design step, we prepare an ensemble of probe molecules $ \thi$
which are supposed to recognize a given and fixed target $\sT =
(\sT_1,\ldots,\sT_N)$. For every possible configuration $\thi$ of the
probe molecule we therefore assign a conditional probability of its
occurrence $\PD(\thi|\sT)$. We describe the conditions of the system by
the Lagrange multiplier $\bD \geq 0$ and demand a canonical Boltzmann
distribution
\begin{equation}
 \PD (\thi|\sT)  =  \frac{1}{Z_\text{D}}  \sum_{\{S\}}
\exp \left[ - \bD \Ham(\sT,\thi;S) \right] , \label{PD} 
\end{equation} 
where the partition function $Z_\text{D}$ guarantees the normalization
$\sum_{\{\thi\}} \PD (\thi | \sT) = 1 $. The sum in (\ref{PD}) extends
over all $2^N$ possible configurations of $S$.  This design step has
been introduced to mimic the process of evolution in nature or design
in biotechnological applications. We remark that the parameter $\bD$,
which can be interpreted formally as an inverse temperature in our
simplifying view of evolution or biotechnological design, basically
controls the degree of optimization of the probe with respect to the
target \cite{Behringer2007}. As recognizing biomolecules are usually
well optimized to each other we typically choose a fairly large value
for $\bD$.

In the second step of our approach, we test the recognition ability of
the designed ensemble. To this end, we consider two copies of the
ensemble of probe molecules at the inverse temperature $\be \geq 0$:
One ensemble is given the target molecule $\sT$, the other system
interacts with a competitive rival molecule $\sR =
(\sR_1,\ldots,\sR_N)$. At that point, we simulate that the probe
molecules have to find their right partner and must decide between the
formation of a complex with the target or the rival. Our aim is to
calculate the free energies of the two possible protein complexes, and
the lower one is then realized in nature. First we evaluate the free
energy for the complex consisting of target or rival and a fixed probe
molecule $\thi$: \beq F(\thi|\sal) & = & - \frac{1}{\be} \ln
\sum_{\{S\}} \exp \left[ -\be \Ham(\sal,\thi;S) \right] ,
\label{Fthisal} \enq for $\al \in \{\text{T $\equiv$ target}, \text{R
  $\equiv$ rival}\}$.  Afterwards we average over the ensemble of
probe molecules using the conditional probability from the design
step and get 
\begin{equation} 
\Fal  =  \sum_{\{\thi\}} F(\thi|\sal)  \PD
(\thi | \sT) .  \label{Fal} 
\end{equation} 
For further investigations we
consider the difference of the free energy $\De F(\sT,\sR) = \FT -
F^\text{(R)}$ as a measure for the selectivity of the recognition
process. For $\De F(\sT,\sR) < 0 \Leftrightarrow \FT < F^\text{(R)}$
the target is recognized by the probe molecules. 

Since we have decided to describe molecular recognition on a very
coarse-grained level it is quite natural that we will also average the
difference in the free energy $\De F(\sT,\sR)$ over all possible
structures of target and rival molecules. Assuming a uniform
probability distribution for both the target's and the rival's
structure, one receives a result $\langle \De F \rangle$ that does not
depend on specific configurations any more. This number can be
interpreted as a characteristic selectivity of the model and its
associated Hamiltonian.

Let us end this section with a brief comment on the restriction of the
contact variable $S_i$ to two distinct values. At first glance, it
might seem that the distinction between only good and bad contacts is
too simple and naive. So one could suggest to consider a finite number
of discrete levels that interpolate between the extreme case of a good
and a bad contact.  This modification accounts for the fact that there
is usually a considerable number of possible alignments between the
corresponding polar moments of opposite amino acids, for example.  It
turns out that the selectivity depends in general on the structural
information contained in the variables $\sT$ and $\sR$. However,
different models of the contact variable do not change the
corresponding functional dependence, although coefficients might be
altered. Thus qualitative conclusions about the behavior of the
selectivity remain the same.  Therefore the simplifying reduction to
two different states of the quality of a contact suffices to describe
molecular recognition in the context of the presented approach. One
can show that the result of $\De F(\sT,\sR)$ is even the same for non
uniformly distributed (discrete or continuous) contact variables, as
long as the distribution is symmetric with respect to the value lying
in the middle between the values for good and bad contacts
\cite{Taktikos2008}.

\section{\label{sec:BadContacts}  Unspecific inclusion of interface water \protect}

In this paper we are mainly concerned with the effect of {\em
  imbedded} solvent molecules at the interface of protein-protein
complexes on the selectivity of molecular recognition. Our basic
approach is based on the hydrophobic effect where bulk solvent degrees
of freedom are already integrated out. The residual solvent degrees of
freedom that show up at the interface as an active part have to be
modeled explicitly. In our approach a solvent molecule can be
imbedded at a position where a bad contact appears. Hence the contact
variable $S_i$ describes the appearance of cavities at the interface.
In this section we will relate the emergence of cavities to thermal
fluctuations. In the next section cavities will be modeled as an
intrinsic geometric feature that is not liable to thermal fluctuations
so that their number is fixed.

Let us allow for an existing cavity to be always filled by a water
molecule that interacts somehow unspecifically with the adjacent amino
acids so that the energy contribution does not distinguish between the
types of the amino acids. This might be interpreted as a van der Waals
contribution which has to be distinguished from a hydrogen bond that
requires certain geometrical and structural prerequisites.  To account
for the geometrical conditions we consider favorable ($- \ga < 0$) or
unfavorable ($\ga > 0$) energy contributions and thus introduce the
variable $w = (w_1,\ldots,w_N)$ with $w_i \in \{-1,1\}$ for the
solvent degree of freedom to distinguish between a favorable ($w_i =
1$) and an unfavorable ($w_i = -1$) energy contribution. The
Hamiltonian then consists of a sum due to the direct contacts at the
proteins' interface as modeled in (\ref{HP}) and a second term due to
the burial of water molecules at bad contact sites:
\begin{equation} 
\Ham(\si,\thi;S,w) := - \eps \sum_{i=1}^N \frac{1+S_i}{2}  
\si_i  \thi_i - \ga \sum_{i=1}^N \frac{1-S_i}{2}  w_i   . \label{HP-S-f}   
\end{equation}
Consistent with observations (e.g. \cite{Covell1997,Levy2006,Li2007})
we request the ratio $\eps/\ga$ to be typically of the order of two to
three.  For a good contact at site $j$ so that water molecules cannot
be imbedded, the variable $w_j$ corresponds to a water molecule of the
bulk and delivers an entropic contribution as it appears in the
summation for the partition function but provides no energy
contribution.

\begin{figure}[tbp]
  \centering \includegraphics[width=0.35\textwidth]{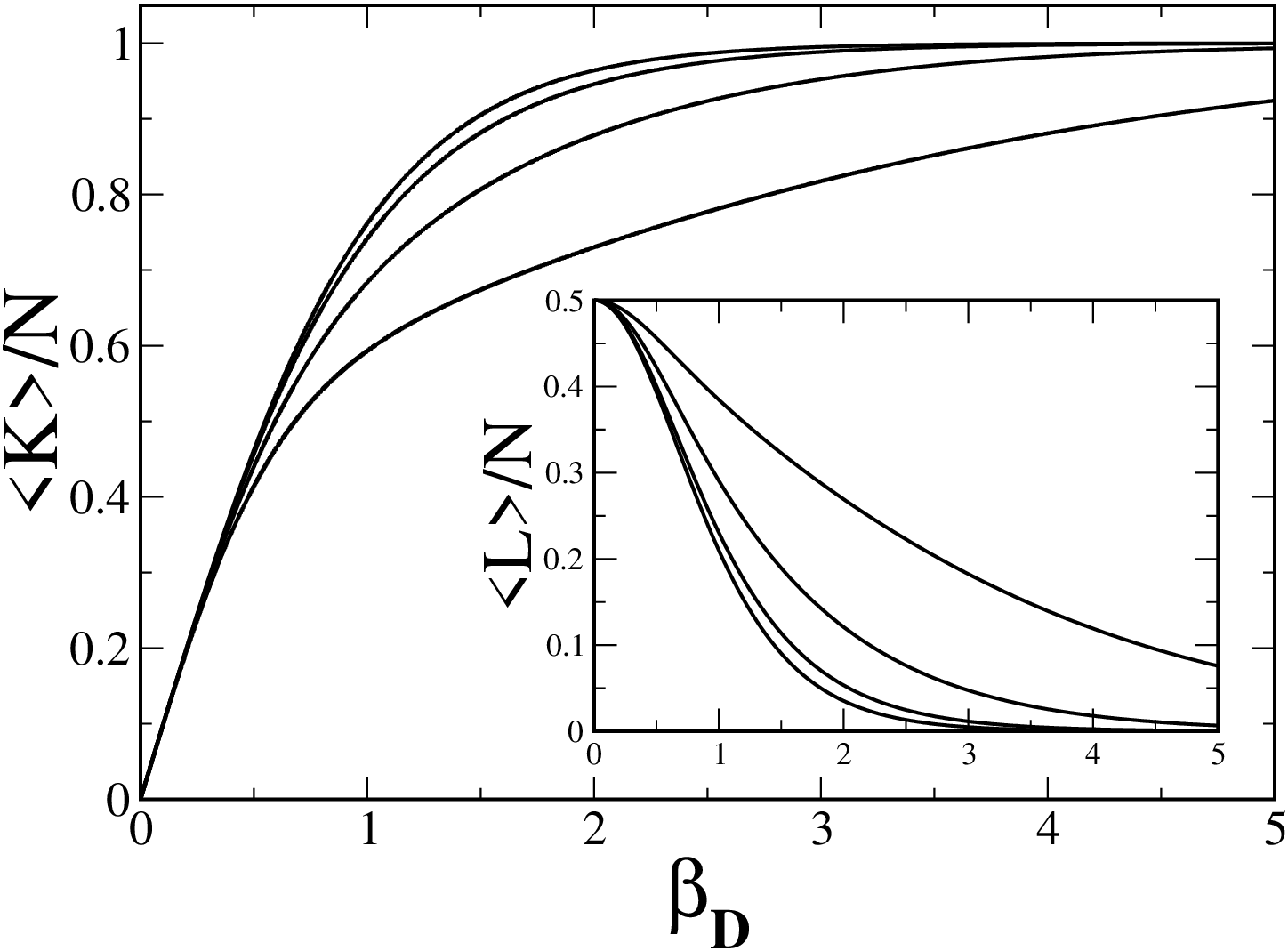}
\caption{Averaged complementarity $\frac{1}{N} \langle K
  \rangle$ as a function of $\bD$ for $\eps =2$.  The energy
  parameter $\ga$ takes the values 0, 0.5, 1, 1.5 (from the left to
  the right). Inset: Normalized number of cavities $\frac{1}{N} \langle L
  \rangle$ as a function
  of $\bD$ for the same parameters ($\ga$ increases from the left to
  the right).}
\label{fig:K-AlsFunktionVonbD}
\end{figure}

We start with the design of the probe ensemble. The calculation of the
conditional probability $\PD(\thi | \sT) = \frac{1}{\ZD} \sum_{\{S\}}
\sum_{\{w\}} \exp\{ -\bD \Ham(\sT,\thi;S,w) \}$ gives
\begin{equation}
\PD(\thi | \sT) = \frac{\prod_{i=1}^N \left[ \exp( \bD \eps \sT_i
    \thi_i ) + 
\cosh(\bD \ga) \right]}{\left[ 4 \cosh \left( \frac{\bD}{2} (\eps+\ga)
    \right) 
 \cosh \left( \frac{\bD}{2} (\eps-\ga) \right) \right]^N} .  \label{PD-Wasser}
\end{equation}
Before giving the result for the difference in the free energy we want
to have a look at some observables of the system which characterize
the design step. We define the complementarity $K$ of the target $\sT$
and a certain probe molecule $\thi$ as $K = \sum_{i=1}^N \sT_i
\thi_i$, whose possible values range from $-N$ to $N$. A value of $K$
close to the maximum $N$ means a high structural complementarity so
that we expect the formation of a complex between target and probe to
become energetically favorable.  We can convert the probability
(\ref{PD-Wasser}) into a distribution for the complementarity
according to $\PD(K) = \sum_{\{\thi\}} \PD(\thi|\sT)
\delta_{K,\sum_{i=1}^N \sT_i \thi_i}$ Using that result to calculate
an averaged complementarity of the designed structures $\thi$ (for
fixed target $\sT$) according to $\langle K \rangle = \sum_{K=-N}^N K
\, \PD(K)$, we finally arrive at
\begin{eqnarray}
\langle K \rangle =  N  \frac{\sinh \left( \frac{\bD \eps}{2} \right)
 \cosh \left( \frac{\bD \eps}{2} \right)}{\cosh \left( \frac{\bD}{2}
 (\eps+\ga) \right) 
 \cosh \left( \frac{\bD}{2} (\eps-\ga) \right)} \label{<K>} . 
\end{eqnarray}
Note in particular that the resulting expression for $\langle K
\rangle$ is independent of the given target $\sT$. Equation
(\ref{<K>}) provides an interpretation for the design parameter $\bD$,
since for large $\bD \rightarrow \infty$ one gets $\langle K \rangle
\rightarrow N$, i.e. the probe molecules are well optimized with
respect to the fixed target and we thus talk of optimal design
conditions.  Further information that we can extract from (\ref{<K>})
concerns the influence of the interaction between the proteins and the
water, given by the parameter $\ga$. As the complementarity is
decreased for increasing $\ga >0$, we can already expect the
selectivity of the recognition to decay as well (compare figure
\ref{fig:K-AlsFunktionVonbD}).

Another observable of interest is the number $L = \frac{1}{2} (N -
\sum_{i=1}^N S_i)$ of cavities at the interface.  Instead of $L$ we
consider the normalized quantity
\begin{eqnarray}
l_{\sT} = \frac{1}{2} \left( 1 - \frac{1}{N}  \sum_{\{\thi\}} \PD(\thi|\sT)  \left\langle \sum_{i=1}^N S_i \right\rangle_{\sT,\thi} \right)  \label{eq11}
\end{eqnarray}
for a certain target $\sT$. The pointed angles $\langle \cdot \rangle$
in (\ref{eq11}) denote a thermal average with respect to the
fluctuating variables $S$ and $w$, the indices indicate that the
structures $\sT$ and $\thi$ are kept fixed.  The result for $l_{\sT}$
proves to be independent of the target's structure and shows also that
the number of cavities increases with increasing $\gamma$ (compare
figure \ref{fig:K-AlsFunktionVonbD}). This has been expected because
for larger $\ga$ there can appear favorable contributions $-\ga$
which first of all require the existence of a sufficient number of
cavities.

\begin{figure}
\centering
\includegraphics[width=0.35\textwidth]{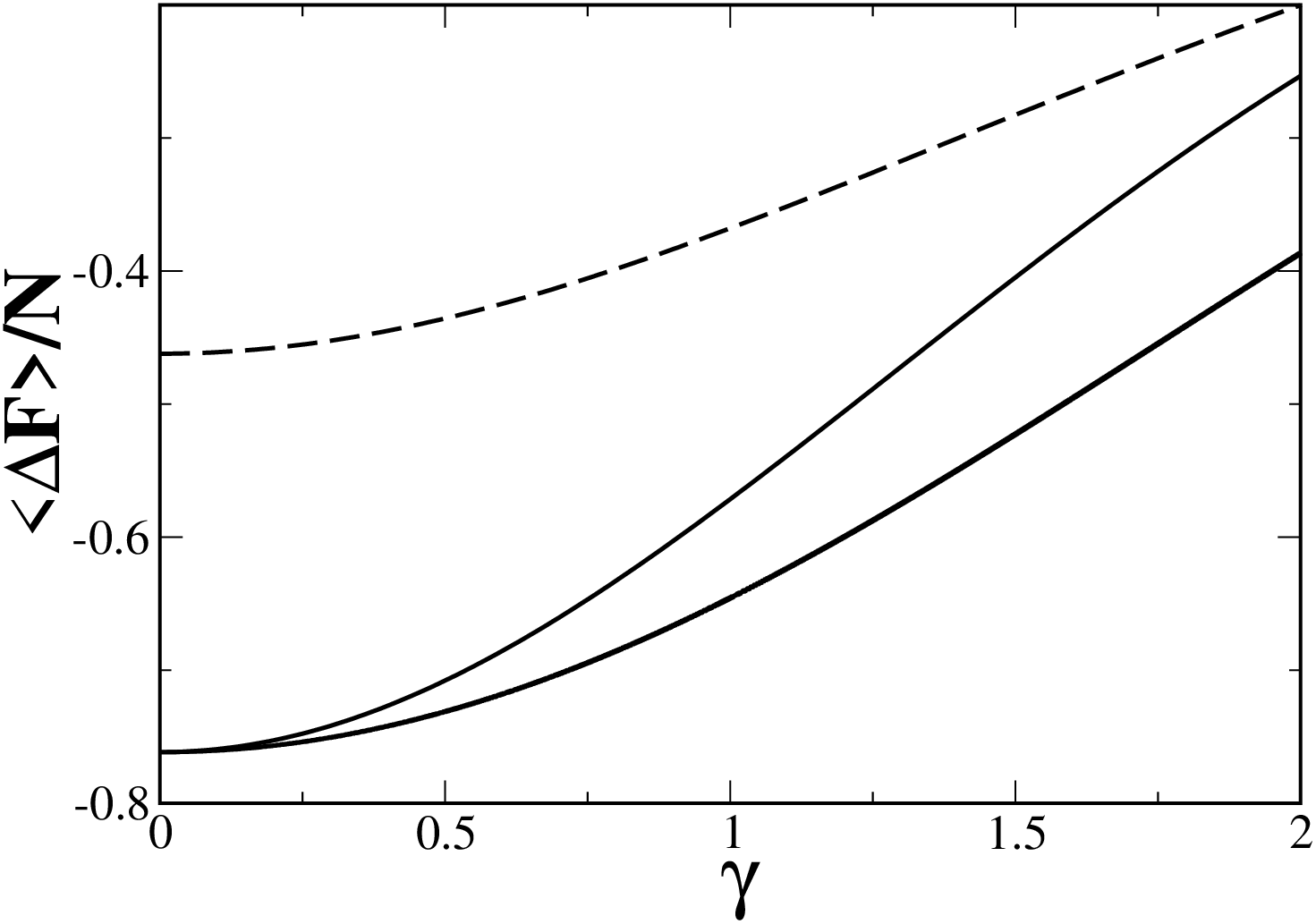}
\caption{
  The averaged selectivity (\ref{sc35}) as
  function of $\ga$ for $\eps=2$ with $\be=1$,
  $\bD=1$ (upper curve) and $\be=0.5$,
  $\bD=1$ (lower curve). The dashed curve corresponds to the
  parameters $\be=1$,
  $\bD=0.5$. An increasing strength of the interaction with the water
  molecules leads to a reduced selectivity.}
\label{fig:AbnahmeSelekt}
\end{figure}

For the analysis of the difference in the free energy of the
interaction with the target and rival, we introduce the function
\begin{equation}
B(\eps,\ga; \be) := 2  +  \frac{1}{\be \eps} \, \ln \left( \frac{1 +
    \exp(- \be \eps) \cosh(\be \ga)}{1 + \exp(\be \eps) \cosh(\be
    \ga)} \right)
\end{equation} 
and obtain the simple result
\begin{equation}
\langle\De F\rangle = - \frac{\eps}{2 N} {\langle K \rangle}(\eps,\ga;\bD)  B(\eps,\ga;\be)  \label{sc35}
\end{equation} 
for the selectivity averaged over all possible target and rival
structures.  Note that $\frac{1}{N} \langle K \rangle (\eps,\ga=0;\bD)
= \tanh \left( \frac{\bD \eps}{2} \right)$ and $B(\eps,\ga=0;\be) =
1$. We compare the characteristic selectivity $\langle \De F \rangle$
of this model with the unmodified case $\ga =0$ and realize that the
selectivity decreases for increasing
values of $\ga$ as shown in figure \ref{fig:AbnahmeSelekt}. To get a
rough estimate of this reduction consider typical values of the
interaction parameters.  We assume a high degree of optimization
during the design step and hence choose $\bD$ to be typically larger
than $\be$. For the selectivity shown in figure
\ref{fig:AbnahmeSelekt} we have chosen the parameters $\eps = 2$, $\be
= 0.5$, $\bD = 1$ and find that the selectivity is then reduced by
15\% for $\gamma = 1$.

The burial of solvent molecules as modeled according to (\ref{HP-S-f})
thus does not lead to an enhancement of the selectivity.  The primary
reason for this is the thermally fluctuating number of cavities so
that for increasing $\gamma$ the system tends to exhibit a larger
number of cavities so that beneficial direct contacts are reduced in
the contribution to the selectivity. Only energy contributions of
direct contacts, however, can discriminate between the differences in
the structures of the recognition sites of the target and the rival.
The energy contributions from imbedded solvent molecules are
insensitive to those differences. For large $\gamma$ the free energy
for the interaction of the probe with the rival becomes more similar
to the one from the interaction with the target and hence selectivity
is reduced. We will come back to this point at the end of section
\ref{sec:enhanced}.

\section{\label{sec:Option} Optimal hydration of geometric cavities  \protect}

In contrast to the previous model, we will now consider protein
interfaces where the number of cavities is an intrinsic geometric
constraint. Cavities appear in the interface as the roughness of the
surface of the proteins might prevent a perfect fit of the shapes of
the two proteins at some positions of the interface. For rigid
proteins the roughness cannot relax and thus one expects the
appearance of a certain number of cavities irrespective of thermal
fluctuations.  Technically, the number of cavities is controlled by a
Lagrange multiplier in our model. So the structure of the molecule is
specified not only by the distribution of amino acids but in addition
by a Lagrange parameter that contains information about the geometry
of the cavities. In addition we allow the cavities to not necessarily
be occupied with water molecules, i.e.  a single cavity can, but does
not have to be filled by solvent. We want to answer the question
whether or not there exists a characteristic fraction of occupied
cavities which leads to the maximum selectivity in the recognition
process.  This enables to distinguish between wet and dry interfaces,
as presented in \cite{Janin1999,Wodak2003,Rodier2005}.

We want to consider a situation where the imbedded water molecules
mediate interactions between the adjacent amino acids. We therefore
require the interaction of a water molecule to depend on the polarity
or hydrophobicity of the adjacent monomers and therefore introduce
three different energy parameters $\gaPP > \gaHP > \gaHH$. Here the
parameter $\gaPP$ specifies the strength of the water-bridged
interaction in a cavity with two adjacent polar residues (PP-cavity),
the parameters $\gaHP$ and $\gaHH$ correspondingly the strength for HP
and HH-cavities. The order of these parameters reflects the fact that
water itself is polar and therefore the interaction with polar
residues is more favorable. Besides, the new parameters have to be
chosen in such a way, that the interaction strength of direct contacts
$\eps$ stays larger. The energetics of the mediated interactions are
intended to mimic hydrogen bonds between the amino acids that are
bridged by solvent molecules. Note that in real interfaces these
bridged hydrogen bonds can involve more than one water molecule
\cite{Hubbard1994,Sonavane2008}. We will, however, only distinguish
between filled and empty cavities, irrespective of the number of
contained water molecules.

Let us now define the $N$-dimensional vector $f=(f_1,\ldots,f_N)$,
whose $i$-th component specifies whether a cavity at site $i$ is
filled by a water molecule ($f_i=1$) or not ($f_i=0$). As we want to
consider interfaces with a fixed total number of cavities we adjust
this number by a Lagrange parameter $\mu$. In addition we consider the
selectivity for varying numbers of  imbedded water molecules
and thus  control the number of filled cavities technically by an additional Lagrange
parameter $\xi$. With the use of
the abbreviations $\al := \gaPP - \gaHP$, $\om := \gaHH-\gaHP$ and
$\eta := \gaHP + \xi$ the additional terms in the Hamiltonian that are
related to the cavities are then
given by
\begin{eqnarray}
 \Ham_\text{cav} & = &  - \sum_{i=1}^N \frac{1-S_i}{2}  f_i \left[ \al \dsim \dthim + \om \dsip \dthip + \eta \right]\nonumber\\
&&  - \mu \sum_{i=1}^N S_i.
   \label{Ham}
\end{eqnarray}
The contact variable $S_i$ thus models the appearance of real
cavities.  Apart form these contributions from solvent in cavities the
total energy of the interface contains the usual contact
Hamiltonian $\Ham_\text{cont}$ as modeled in (\ref{HP}) so that $\Ham
= \Ham_\text{cont} + \Ham_\text{cav}$.

The strategy to calculate the selectivity for the above discussed
model is outlined in the appendix.  The Lagrange parameters are used
to fix the (normalized) number $l$ of cavities in the interface and
the fraction $f$ of cavities that are filled with water. We will
utilize the normalization that $f \in [0,l]$. We thus obtain the
selectivity $\left\langle \De F \right\rangle_l (f)$ for interfaces
with a fixed number of cavities as a function of the number of
imbedded molecules.  We note that the actual results, that are
presented in the subsequent subsections, are obtained with a
\texttt{Mathematica} program.

\subsection{\label{sec:enhanced}Selectivity enhancement }

As we want to compare protein interfaces with imbedded water
molecules, with the dry realization ($f=0$), we consider the
correction factor $ C_l(f) := \frac{\left\langle \De F \right\rangle_l
  (f)}{\left\langle \De F \right\rangle_l (f=0)}$. The range over $f$
with $C_l(f) > 1$ corresponds to increased selectivity of molecular
recognition, whereas a correction factor with $C_l(f) < 1$ describes
lowered selectivity due to the inclusion of solvent molecules.  Now we
are interested in the probability of the macroscopic realization for a
wet interface, described by the parameters $l$ and $f$, in contrast to
a dry interface and obtain as a rough estimate
\begin{eqnarray}
\frac{\text{Prob}^\text{(with water)}}{\text{Prob}^\text{(dry
    interface)}} \approx  
\frac{e^{- \be \langle \De F_l(f) \rangle}}{e^{- \be \langle \De
    F_l(f=0) \rangle}}  \approx e^{N (C_l(f)-1)}.    \label{estim}
\end{eqnarray}

\begin{table}
        \centering
                \begin{tabular}{lllll}
                        A: $\qquad$ &  $\eps=2 \qquad$ & $\gaPP=1 \quad$ & $\gaHH=-1 \quad $    & $\gaHP=0.5$     \\
      B: $\qquad$ &  $\eps=2 \qquad$ & $\gaPP=1 \quad$ & $\gaHH=-0.5 \quad $  & $\gaHP=0.5$    \\
      C: $\qquad$ &  $\eps=2 \qquad$ & $\gaPP=1 \quad$ & $\gaHH=-0.5 \quad $  & $\gaHP=0$ 
                \end{tabular}
\caption{Investigated sets of energy parameters in (\ref{Ham}).}
\label{werte}
\end{table}

To obtain an impression of the size of a possible enhancement of the
selectivity due to the inclusion of water we have to choose a
characteristic set of the involved parameters. For the discussion we
will consider interfaces whose fraction of cavities varies from 10\%
to 30\% ($l=0.1\ldots0.3$) which seems to be reasonable for natural
protein-protein interfaces \cite{Hubbard1994,Sonavane2008}. In the
following we will discuss the results for $l=0.3$ and note that for
$l=0.1$ and $l=0.2$ we obtain qualitatively similar results.
Concerning the energy parameters $\eps$, $\gaPP$, $\gaHP$ and $\gaHH$,
we will consider exemplarily three different combinations, denoted by
A, B, and C as shown in table \ref{werte}.  For all combinations of
parameters the inclusion of a water molecule in a PP-cavity is
energetically most favorable, whereas the interaction of a water
molecule with at least one polar residue in a HP-cavity is more
beneficial than in a purely hydrophobic HH-cavity.  Going from A to B
we leave $\eps$, $\gaPP$ and $\gaHP$ unchanged, while the change of
the parameter $\gaHH$ from $-1$ to $-0.5$ reduces the penalty for an
inclusion of water between two hydrophobic residues. Accordingly, at
the change from B to C, the occupation of water between different
types of amino acids becomes less favorable. Furthermore, we set
$\bD=1$ as we want to have a high degree of optimization during the
design and $\be=0.5$, satisfying the relation $\be \eps = \Ord (1)$.

\begin{figure}
        \centering
               \includegraphics[width=0.35\textwidth]{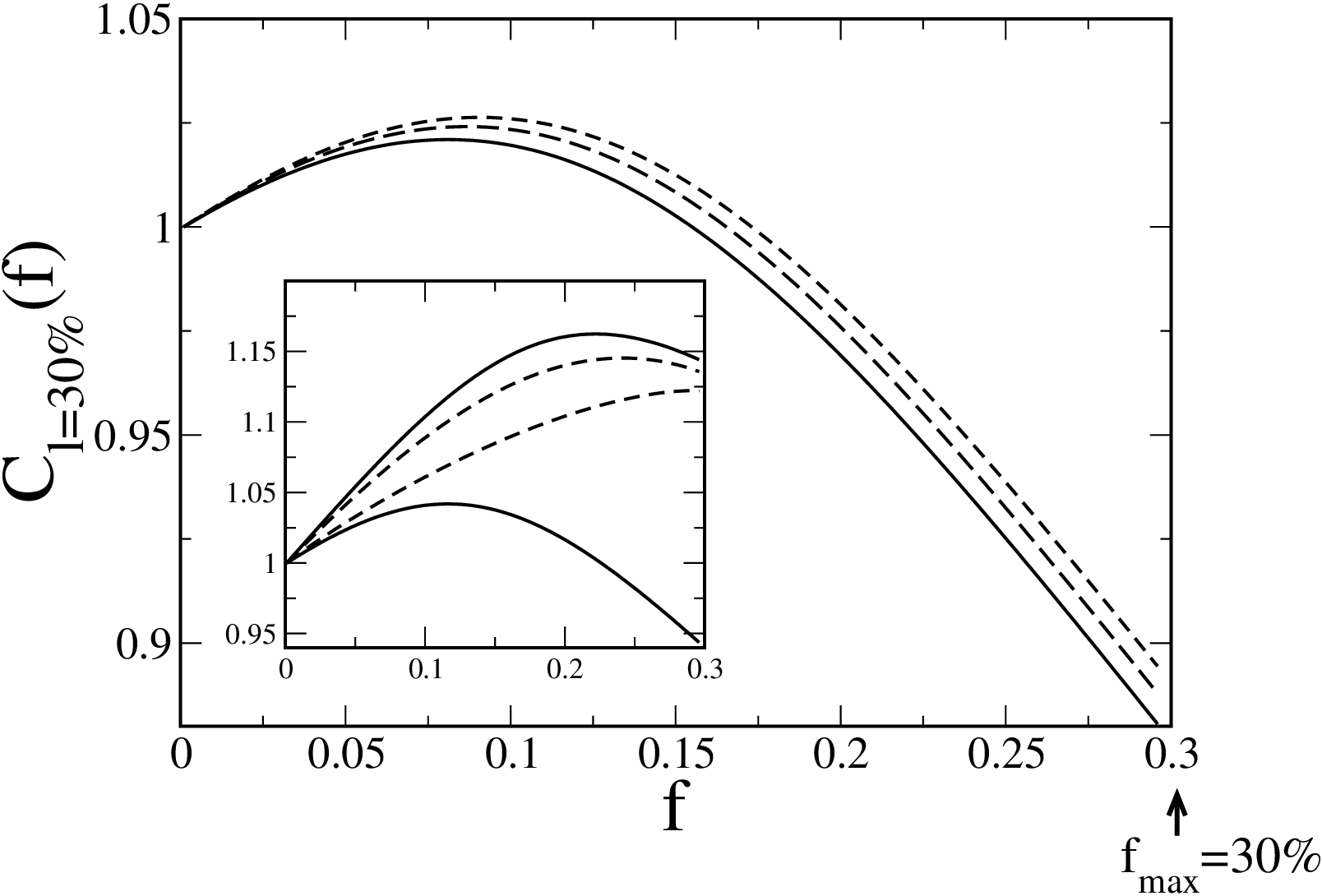}
        \caption{Analysis of $C_{l=30\%}(f)$ for the parameter set A
               and $\bD=1$, $\be=0.5$, $N=32$ (the inset shows
               parameter set B (lower curve) and C (upper curve), the
               dashed curve corresponds to set C for $\bD=0.8$ and 0.6
               from above). The exactly averaged correction factor
               is shown together with the approximation discussed in
               the appendix ($N=32$ and 64 from above). The maximum at $\fopt \approx0.085$
               leads to an enhancement factor of 1.9 (compare relation
               (\ref{estim})). Note that for the exact average
               (\ref{DeF(f)}) a value for $N$ has to be
               specified. Different choices, however,  show only very small finite-size variations. }
        \label{fig:neuA-s04}
\end{figure}

The correction factor $C_{l=30\%}(f)$ for parameter set A and an
interface with 30\% cavity area is plotted in figure
\ref{fig:neuA-s04} for $N=32$.  We were able to show in general, that
the correction factor $C_l(f)$ features a characteristic maximum for
some value of $f$, say $\fopt$, with $C_l(\fopt) >1$ which is lying
somewhere in the allowed interval of $f$. The existence of a maximum
for $C_l(f)$ means that there is a fraction of occupied cavities for
which the selectivity of the recognition process becomes maximum. For
the considered parameters this typically results in an enhancement of
the selectivity for a hydrated interface by a factor of two to four
using the estimate (\ref{estim}) with $N=32$. Note that $N \sim \Ord
(30)$ holds for typical interfaces in natural protein-protein
complexes \cite{Wodak2003,Janin2007}. The presented example shows
that, for a interface with 30\% cavities roughly one third of the
cavities should be filled with water molecules on average to give
maximum selectivity.  Interestingly, the selectivity first raises up
to a maximum with $C_l(\fopt)>1$ and afterwards gets even smaller than
one.

We now want to obtain a  physical understanding for the
evolution of the correction factor $C_l(f)$ which can show both an
enhancement and a reduction of the selectivity depending on the degree
of hydration of the interface.
 To this end we consider various observables
which characterize the interface between the probe molecule and the
target molecule in more detail. The observables provide
an answer to the question between which pairs of residues
the cavities or the direct contacts are distributed. We define
the following quantities which are averaged over the ensemble of probe
molecules:
\begin{eqnarray}
\WPPnHT  = \sum_{\{\thi\}}  \WPPsTth  \PD(\thi|\sT)  , 
\end{eqnarray}           
\begin{eqnarray}
\WHHnHT  = \sum_{\{\thi\}}  \WHHsTth  \PD(\thi|\sT)  , 
\end{eqnarray}           
and
\begin{eqnarray}
\WHPnHT  = \sum_{\{\thi\}}  \WHPsTth  \PD(\thi|\sT)  . 
\end{eqnarray}           
These quantities specify how often a particular type of cavity is
realized in the interface.  We have chosen the index $\nHT =  \NHT/N$ because
the obtained expressions only  depend on the target's
hydrophobicity $\NHT  = \sum_{i=1}^N \dTp$.  The formula for $\WPPsTth$ is given by
\begin{eqnarray}
\WPPsTth  : =  \frac{1}{N} \left\langle \sum_{i=1}^N \frac{1-S_i}{2}  f_i  \dTm \dthim \right\rangle_{\sT,\thi}   \label{PPsTth} 
\end{eqnarray}
for fixed target $\sT$ and fixed probe molecule $\thi$, similar
definitions hold for $\WHHsTth$ and $\WHPsTth$. The corresponding
functions for the direct contacts are indicated by the letter $D$:
\begin{eqnarray}
\DPPoWnHT   =  \sum_{\{\thi\}}  \DPPoWsTth  \PD(\thi|\sT) ,
\end{eqnarray}           
\begin{eqnarray}
\DHHoWnHT   =  \sum_{\{\thi\}}  \DHHoWsTth  \PD(\thi|\sT) ,
\end{eqnarray}           
and
\begin{eqnarray}
\DHPoWnHT   =  \sum_{\{\thi\}}  \DHPoWsTth  \PD(\thi|\sT) . \label{op44}
\end{eqnarray}           
The definition of $\DPPoWsTth$ is analogue to the previous functions
depending on a fixed target and probe molecule:
\begin{eqnarray}
\DPPoWsTth   :=  \frac{1}{N} \left\langle \sum_{i=1}^N \frac{1+S_i}{2}  \dTm \dthim \right\rangle_{\sT,\thi}  .   \label{PPoWsTth}
\end{eqnarray}

\begin{figure}
\centering
\includegraphics[width=0.35\textwidth]{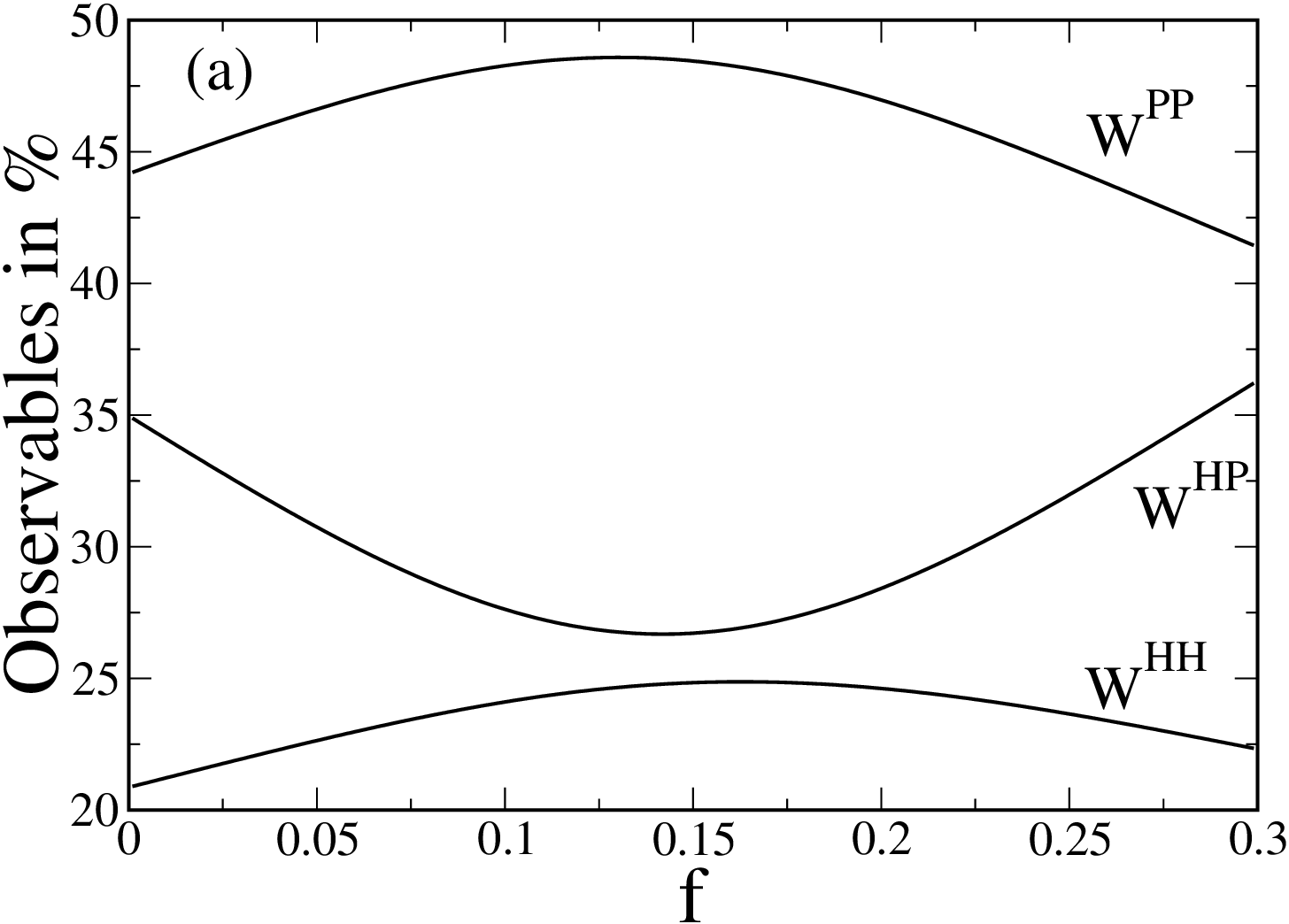}
\includegraphics[width=0.35\textwidth]{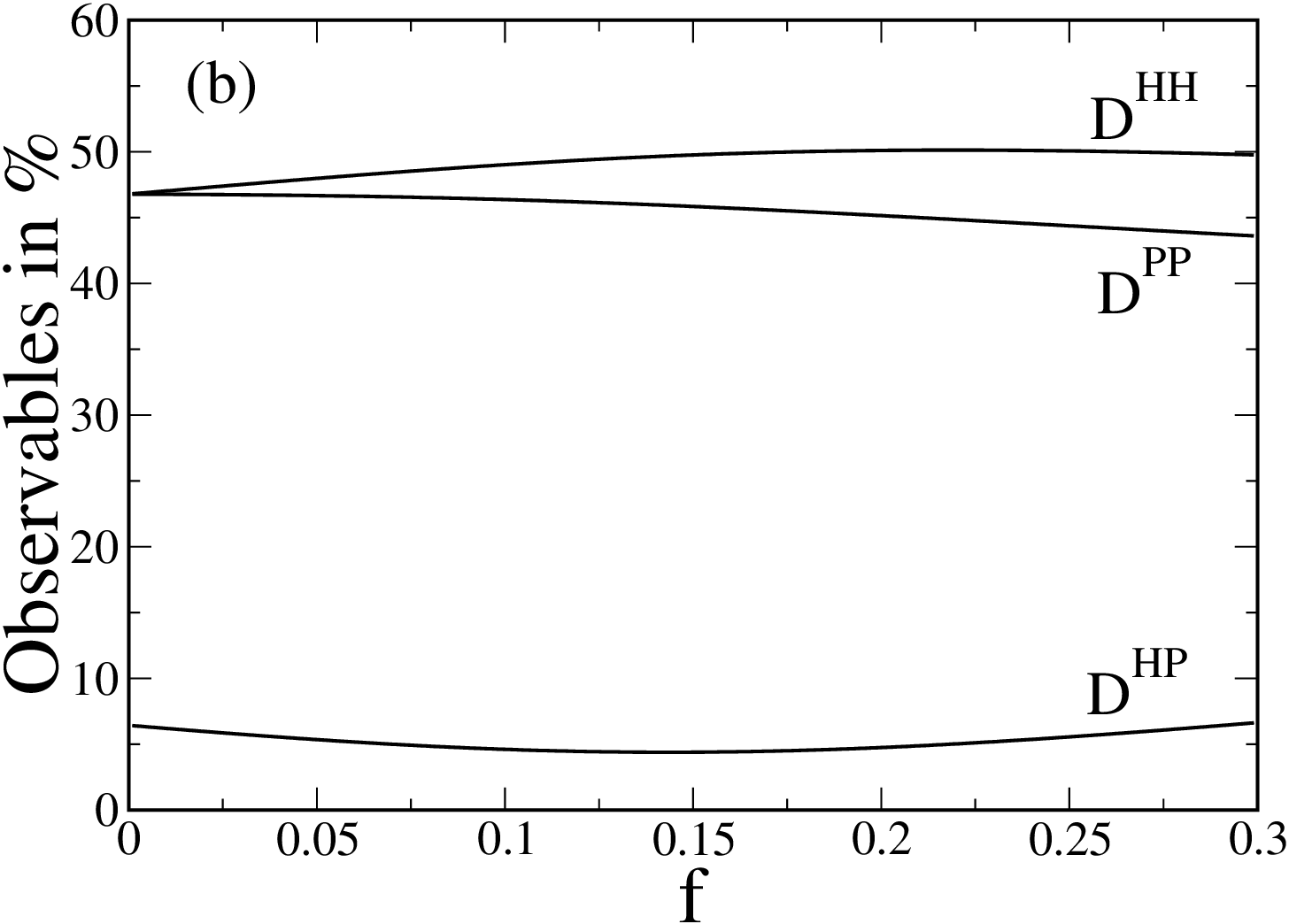}
\caption{Analysis of the correction factor $C_{l=30\%}(f)$ (shown in
  figure \ref{fig:neuA-s04}) for the parameter set B and
  $\bD=1$, $\be=0.5$, $N=32$.  
  (a) observables of cavities $\WPP$, $\WHH$, $\WHP$; (b) observables of
  direct contacts $\DPP$, $\DHH$, $\DHP$. The observables are
  evaluated for $\nHT=0.5$ and are shown in dependence on the
  normalized $f$ (in \%).}
        \label{fig:neuB-s04}
\end{figure}

In figure \ref{fig:neuB-s04} we have normalized the given observables
to the sum of all direct contacts and to the sum of all occupied
cavities respectively. Since the observables have to be computed for a
certain hydrophobicity of the target, we have chosen the typical value
of $\nHT = 0.5$.  Note that the $\nHT=0.5$ terms in (\ref{DeF(f)})
dominate the sum and hence this also corresponds approximately to an
average over all target structures for sufficiently large $N$ (see
discussion in the appendix). For values of $0 \leq f \leq 0.13$ we see
that the fraction of favorable imbedded water molecules in
PP-cavities increases. For a small fraction of filled cavities the
solvent molecules will preferentially be imbedded in PP-cavities due
to the large energy gain they can provide. This goes along with a weak
decrease of direct PP-contacts. For a further increasing number of
water molecules eventually all PP-cavities will be used up and water
molecules have to go into the HP-cavities as they provide still an
energy gain. The relative fraction of occupied PP-cavities therefore
will be reduced for increasing $f$. This subsequent decrease of the
PP-fraction goes along with a decreasing selectivity of the
recognition process.  We notice that the observables in figure
\ref{fig:neuB-s04} take similar values for $f=0$ and $f=\fmax = l$
though $C_l(0)=1$ is quite different from $C_l(\fmax)$ which can even
be smaller than one.  This demonstrates that the competitive influence
of the rival on the selectivity gets more and more important for an
increasing number of buried water molecules. Note, however, that the
selectivity does not change its sign, so we still have recognition of
the target by the probe molecules.

Looking at the observables that describe direct contacts (see figure
\ref{fig:neuB-s04}), we observe that they show only a weak dependence
on $f$. The fraction of $\DPP$ and $\DHH$ strongly dominate the direct
contacts between different types of amino acids. For $f>0$ we get
$\DHH > \DPP$ which can be explained in the following way: For an
existing site with opposite polar residues (PP) it is more beneficial
to fill a cavity with water (in comparison to a HH-cavity), and
therefore the HH-sites are more likely used for direct contacts
between the amino acids.

For all results shown in this subsection a high degree of optimization
has been assumed ($\bD = 1$ in comparison to $\be = 0.5$). If the
quality of the design is reduced by decreasing the parameter $\bD$ the
observed effect of an enhancement of the selectivity due to the
inclusion of water molecules in the interface is still present, but
becomes weaker and weaker (see inset of figure \ref{fig:neuA-s04}).
Even for a situation with $\beta = \bD = 0.5$ selectivity enhancement
due to hydration can appear although we note that this is not the case
for all sets of the $\gaHH$, $\gaPP$ and $\gaHP$ parameters (namely
only for set C of the three considered ones).

\begin{figure}
\centering
\includegraphics[width=0.35\textwidth]{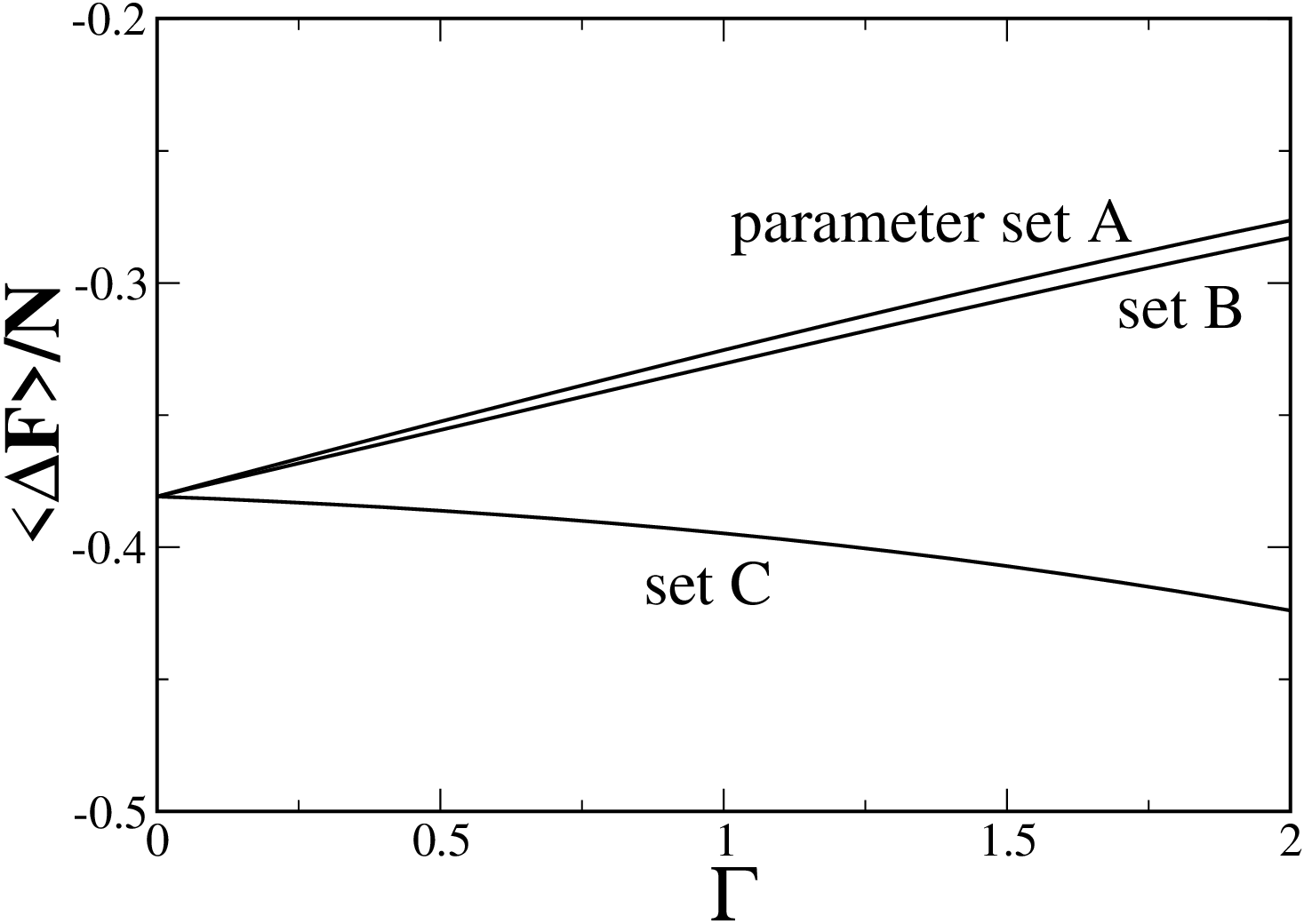}
\caption{
  The averaged selectivity for the model (\ref{HamGa}) with a thermally
  fluctuating number of cavities as
  function of $\Gamma$ for $\be=0.5$,
  $\bD=1$ and the different relative adjustments of the parameters
  $\gaPP$, $\gaHH$ and $\gaHP$ as specified in table \ref{werte}.}
\label{fig:AbnahmeSelektGrossGamma}
\end{figure}

As a final comment let us come back to the situation where the
appearance of cavities is due to thermal fluctuations and where different to
the considerations in section \ref{sec:BadContacts} the energy
contributions from embedded water particles now distinguish between
the different types of amino acids of the cavities. This is
technically incorporated if the Lagrange parameters in (\ref{Ham}) are
set to zero and thus the number of cavities and the number of imbedded
solvent molecules fluctuate. The cavity part of the Hamiltonian reads
\begin{equation}
 \Ham_\text{cav}  =   - \sum_{i=1}^N \frac{1-S_i}{2}  f_i \Gamma
 \left[ \al \dsim \dthim + \om \dsip \dthip + \gaHP \right] \label{HamGa}
\end{equation}
where the parameter $\Gamma$ specifies the relative weight of the direct
contacts and the water-mediated interactions. Notice that different to
(\ref{HP-S-f}) no distinction between a favorable and an unfavorable
energy contribution of an embedded water molecule is incorporated. The
selectivity as a function of the parameter $\Gamma$ is shown in figure
\ref{fig:AbnahmeSelektGrossGamma} (compare also figure
\ref{fig:AbnahmeSelekt}).  One finds that the inclusion of water
molecules might lead to an enhanced selectivity although an
enhancement of the selectivity is not observed for all considered
parameter sets. So the distinction of the type of amino acids which
are participating in water-mediated interactions is crucial for the
appearance of an enhanced selectivity due to hydration. We also
conclude that the appearance of rigid cavities seems to facilitate the
enhancement of selectivity.

\subsection{\label{sec:dry}Dry and wet interfaces}

The last part of our investigation examines the influence of the
 hydrophobicity of the interface on the enhancement of the selectivity of the
recognition process.  In the previous subsection an average over all
possible hydrophobicities has been carried out so that the discussed
results are general statements formulated for all classes of proteins
and can be understood as a characteristic property of the considered
model for molecular recognition. In nature, however, the
hydrophobicity is typically different for proteins that fulfill
different biological tasks. For example, the average hydrophobicity of
the interface of antigen-antibody complexes is relatively small
(comparable to the rest of the surface of the protein that is
exposed to bulk water)
whereas the interfaces of enzyme-inhibitor complexes are largely
hydrophobic \cite{Jones1996,Wodak2003,Janin2007}. For this reason, we
are also interested in an analysis of the free energy difference for a
given class of proteins with fixed (averaged) hydrophobicity $\langle \nHT \rangle$ of the
target (see the appendix for the details how the corresponding
correction factor $C_l(\langle \nHT \rangle;f)$ is evaluated).

\begin{figure}[htbp]
\centering
\includegraphics[width=0.35\textwidth]{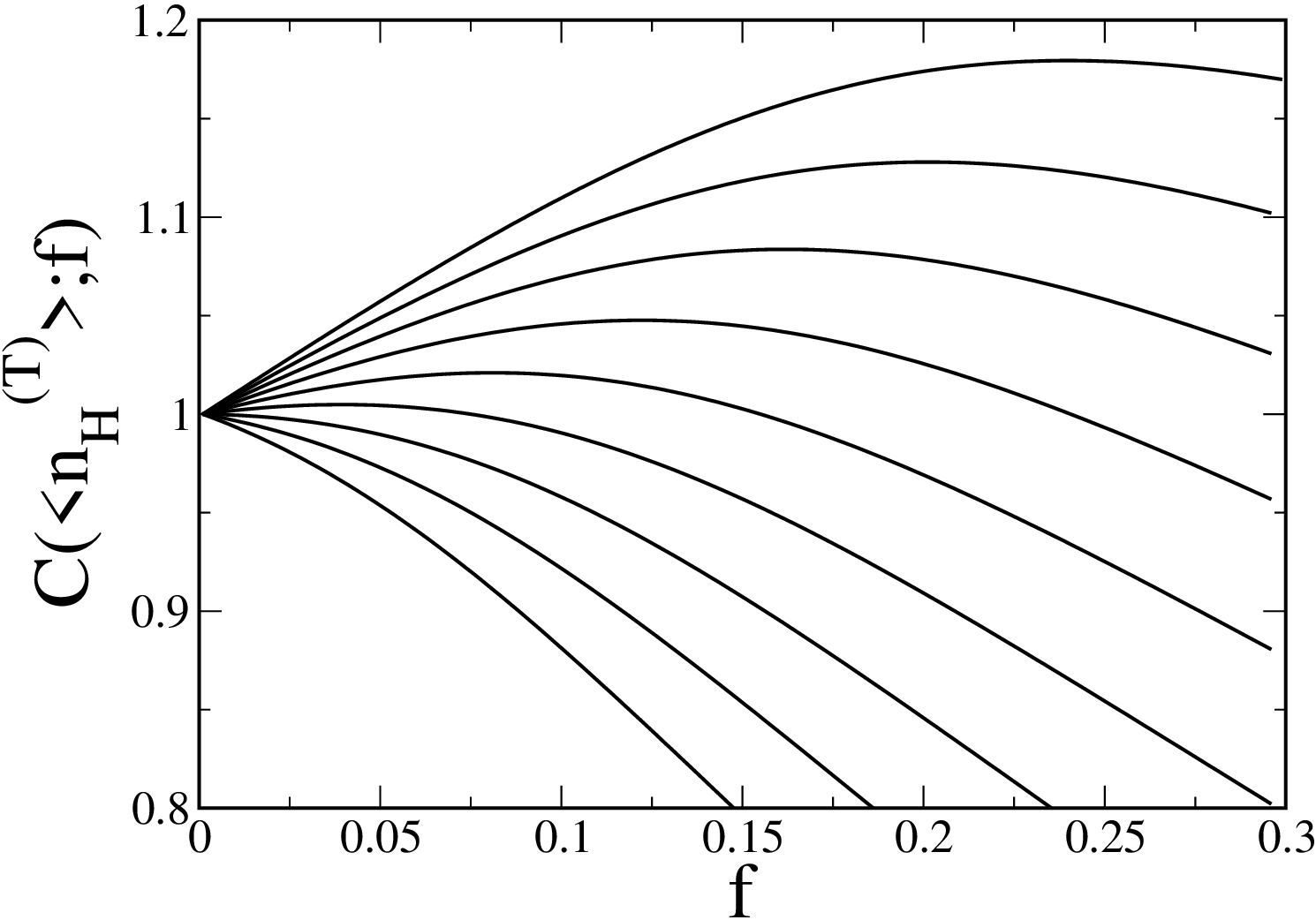}
\caption{Correction factor $C_l(\langle \nHT \rangle;f)$ as a
  function of the fraction $f$ of occupied cavities for different
  fixed hydrophobicities $\langle \nHT \rangle$ ranging from $0.1$ to
  $0.9$ in units of $0.1$ from top to bottom (parameter set A and
  $l=0.3$).  }
\label{fig:ABC-s=04_nht}
\end{figure}

\begin{figure}[htbp]
\centering
\includegraphics[width=0.35\textwidth]{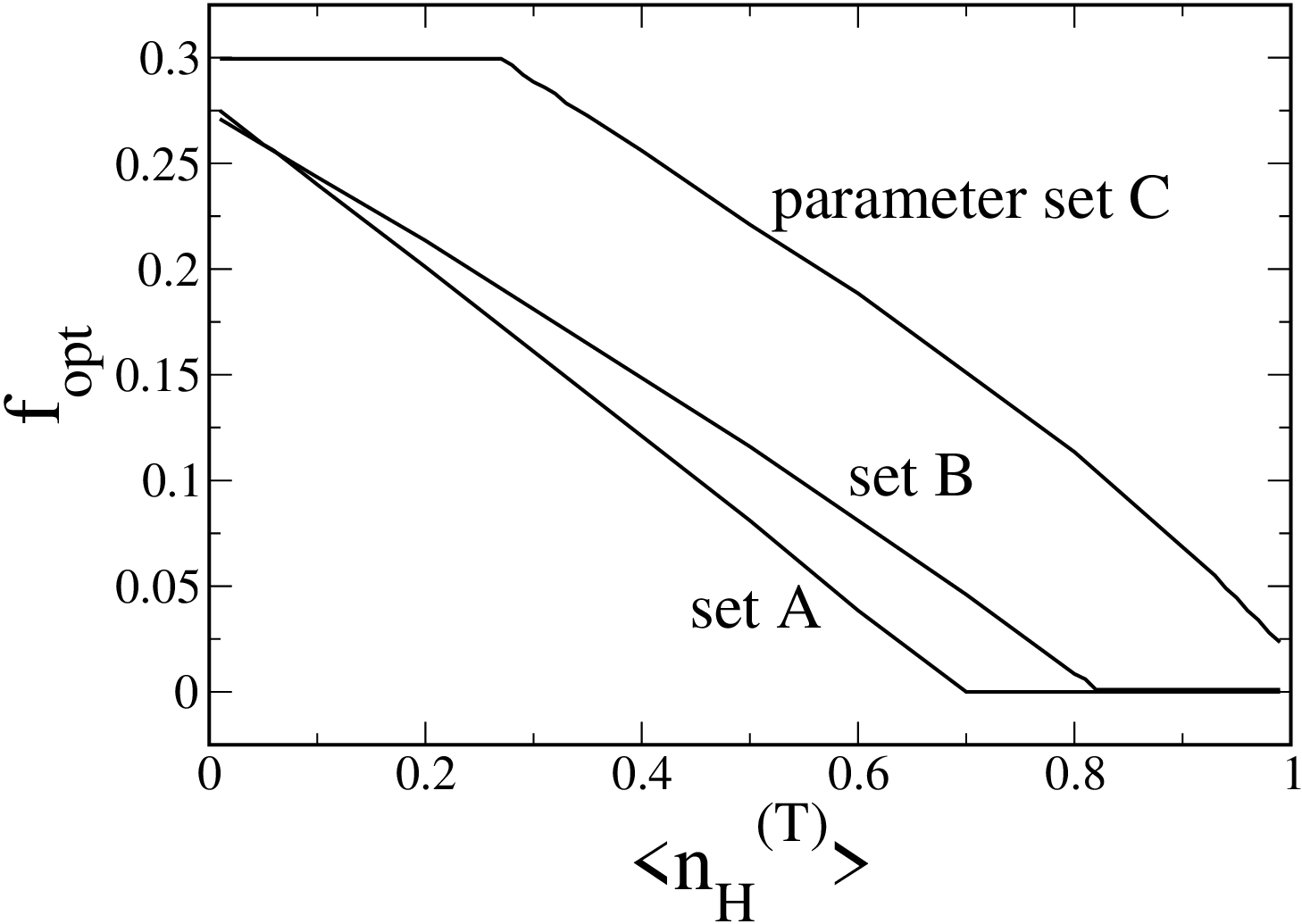}
\caption{Position $\fopt$ of the selectivity maximum as a
  function of the hydrophobicity $\langle \nHT \rangle$ of the
  interface with 30\% cavities. $\fopt = 0$ favors a
  dry interface, $\fopt = 0.3$ corresponds to a maximally
  hydrated (wet) protein interface.}
\label{fig:ABC-s=04}
\end{figure}

We get the result that the correction factor $C_l(\langle \nHT
\rangle;f)$ for given averaged hydrophobicity $\langle \nHT \rangle$
of the target molecules develops a characteristic maximum with
$C_l(\langle \nHT \rangle) >1$ for small hydrophobicities so that the
selectivity is remarkably enhanced in comparison to the complex with a
dry interface (see figure \ref{fig:ABC-s=04_nht}).  For a
protein-protein complex with a given small hydrophobicity of the
interface the scenario of a dry interface is thus less favorable than
the scenario with a hydrated interface. The position $\fopt$ of the
optimum filling fraction for the class of proteins with a fixed
hydrophobicity demands a shift from wet to dry interfaces when the
hydrophobicity is increased as shown in figure \ref{fig:ABC-s=04}. We
note, however, that for complexes with large hydrophobicities the
recognition is still selective. One also observes that the transition
between an optimal dry and wet interface depends on the chosen
parameter values for the coupling constants. Our findings thus
reproduce the empirically found correlation that the degree of
hydration at protein-protein interfaces decreases with the
hydrophobicity of the interface (compare
\cite{Janin1999,Wodak2003,Rodier2005}).

We conclude this subsection by considering the modification of the
model (\ref{Ham}) with no discrimination of the type of cavity with
respect to the energy gain when water is imbedded. In terms of the
coupling parameters this means that we have $\gaPP = \gaHP = \gaHH =
\ga$. Again we observe the appearance of a characteristic optimum
fraction of occupied cavities which maximizes the selectivity such
that $C_l(\langle \nHT \rangle;\fopt) > C_l(\langle \nHT \rangle; f=0)
=1$.  However, if we again consider interfaces with a varying
hydrophobicity $\langle \nHT \rangle$ of the target the position
$\fopt$ of the selectivity maximum is not shifted as can be understood
from the fact that the energy gain due to imbedding water molecules
cannot resolve the hydrophobicity of the interface. Consequently no
transition from a dry to a wet interface shows up for this
modification of the cavity Hamiltonian.

\section{\label{sec:Summary} Summary \protect}

On the basis of coarse-grained modeling we have investigated the
influence of solvent molecules on molecular recognition and found that
they can provide an enhanced selectivity. To describe the molecular
recognition, we have adopted a two-stage approach containing a design
of probe molecules and a testing of their recognition ability. The
energy that stabilizes the protein-protein complex is described in a
coarse-grained view on the level of the hydrophobicity of the amino
acids and the residual solvent molecules imbedded at the interface.

We discussed a model with an inclusion of water molecules in every
cavity at the interface without any coupling to the composition of
residues of the two proteins. For all kinds of additional interaction
strengths the selectivity of the recognition process is then decreased.
The focus of our investigation was then set on the model with an
optional inclusion of water molecules at the interface. Additionally
the interaction of water depends on the adjacent types of monomers.
Having fixed the average number of cavities at the proteins' interface
as an intrinsic geometric constraint we have found that there is a
characteristic fraction of occupied cavities such that the selectivity
becomes maximum. We showed that in many cases it is advantageous to
have an occupied fraction in between 25\% and 75\%. The probability to
have recognition of the correct target molecule is then typically
enhanced by a factor of two to four. In addition we could establish a
correlation between the degree of hydration of the interface and its
hydrophobicity which naturally leads to a discrimination of dry and
wet interfaces. We thus reproduce empirical findings for real
protein-protein interfaces on the level of a coarse-grained model.  We
finally conclude that imbedded solvent molecules have to be considered
as an active part of molecular recognition processes and can
considerably contribute to the selectivity.

\appendix* 

\section{Evaluation of  the selectivity}

In this rather technical appendix we outline the strategy to evaluate
the selectivity of the recognition process where a specified fraction
of cavities is filled with water molecules. The energy contributions
at the interface are modeled by the Hamiltonian (\ref{Ham}).

Following the two-step-approach to obtain the selectivity, we first
calculate the conditional probability $\PD(\thi|\sT) = \frac{1}{\ZD}
\sum_{\{S\}} \sum_{\{f\}} \exp\{ -\bD \Ham(\sT,\thi;S,f) \}$ in the
design step.  We emphasis that the Lagrange parameters $\mu =
\mu(\sT,\thi;\bD)$ and $\xi = \xi(\sT,\thi;\bD)$ that have to be used
for the design both depend on the structure of the target $\sT$ and a
certain probe molecule $\thi$ and the design conditions $\beD$.  For
each interaction of the probe with a molecule (target or rival) a
different set of Lagrange parameters has to be specified in the most
general situation. At the transition to the testing step, we
consequently need to introduce additional sets of different Lagrange
multipliers corresponding to the interaction of the probe with both
the target $\sT$ and the rival molecule $\sR$ at inverse temperature
$\be$.  However, this most general treatment is rather cumbersome.
Instead, we fix the number of given cavities and the fraction of the
occupied cavities at the design step and attribute this adjustment as
an intrinsic geometric property to the probe molecules which is
conserved at the testing step. In doing so only one set of Lagrange
parameters is necessary in the testing step. This set is determined in
the design and exhibits a dependence on the previously fixed structure
of the target.  The structure of the probe molecule at the interface
is thus specified by the set $(\thi, \mu(\sT,\thi), \xi(\sT,\thi))$.
The cavity part of the Hamiltonian for the testing step is hence given
by (\ref{Ham}) with the set of Lagrange parameters obtained in the
design step.

Before we can evaluate the free energy difference we have to calculate the
 Lagrange multipliers $\mu$ and $\xi$. Since the fixing of the
expectation values of the normalized number of cavities $l$ and the
fraction of occupied cavities $f$ suffices to be softly implemented
for the ensemble of probe molecules, we just regard the averaged
quantities $\lsT = \sum_{\{\thi\}}
\lsTth \, \PD(\thi|\sT)$ and $ \fsT  =
\sum_{\{\thi\}} \fsTth \, \PD(\thi|\sT)$ where 
\begin{eqnarray}
\lsTth =  \frac{1}{N}    \sum_{i=1}^N\left\langle \frac{1- S_i}{2} \right\rangle_{\sT,\thi}  
\end{eqnarray}
and 
\begin{eqnarray}
\fsTth = \frac{1}{N} \sum_{i=1}^N  \left\langle  \frac{1-S_i}{2}f_i \right\rangle_{\sT,\thi}  
\end{eqnarray}
denote thermal averages in the design step (including the Lagrange
parameters) with fixed $\sT$ and $\theta$. One can show that the
analytically obtained results for $\lsT$ and $\fsT$ do not depend on
the exact structure of $\sT$ but only on the target's hydrophobicity
$\NHT$ given by $\NHT = N \nHT = \sum_{i=1}^N \dTp$.

The free energy turns out to be determined by the structural
differences between the recognition sites of the target and the rival.
To write the result of the free energy difference in a compact way we
need to define quantities that specify the differences of the target and
the rival molecules. We thus define $X = \sum_{i=1}^N \dTp \dRm \in
\{0,\ldots,\NHT \}$ and $Y = \sum_{i=1}^N \dTm \dRp \in
\{0,\ldots,N-\NHT \}$.  The free energy difference for a given target
and rival structure is then given by
\begin{eqnarray}
\De F(\sT,\sR) = -\frac{1}{\be} \, B(\al,\om) \, X     -\frac{1}{\be} \, B(\om,\al) \, Y  ,  \label{op8}
\end{eqnarray}
where we have introduced the function $B(\al,\om)$
\begin{align}
B(\al,\om) := &  \frac{ G_\text{D}(\eps,\om) \ln
  \frac{G(\varepsilon,\om)}{G(-\eps,0)} + 
G_\text{D}(-\eps,0) \ln \frac{G(-\varepsilon,0)}{G(\eps,\alpha)}} 
 {4 e^{2 \bD \mu} \cosh(\bD \eps) +  2 + e^{\bD \eta} \left( 1 + e^{\bD \om } \right)}
\end{align}
with
\begin{equation}
  G (x,y) = 2 e^{\be x + 2 \be \mu} +1+ e^{\be (\eta + y)}
\end{equation}
and similarly
\begin{equation}
  G_\text{D} (x,y) = 2 e^{\bD x + 2 \bD \mu} +1+ e^{\bD (\eta + y)}.
\end{equation}
Note that the auxiliary function $B(\al,\om)$ implicitly depends on
the structure $\sT$ of the target through the dependency on the
Lagrange parameters. As already mentioned above, this dependence is,
however, reduced to a dependence on the hydrophobicity $\NHT$ of the
target, that is $B(\al,\om)=B(\al,\om; \NHT)$.  For this reason,
averaging over all possible structures of the target and rival
molecules will "only" demand the computation of $\Ord(N)$ terms
instead of an explicit evaluation for all $2^N$ configurations. Using
the expressions for $\lsT = \lsT(\NHT;\muNHT,\xiNHT)$ and $\fsT =
\fsT(\NHT;\muNHT,\xiNHT)$, we can set $\lsT$ and $\fsT$ to some
desired numbers $l$ and $f$, respectively, and get numerically the
values of $\muNHT$ and $\xiNHT$.

Instead of computing $\De F(\sT,\sR)$ for a specific configuration of
$(\sT,\sR)$ --- or due to the sole dependence on the hydrophobicity,
for a given combination $(\NHT,\NHR)$ --- we average over the ensemble
of probe molecules which leads to the expression $\langle \De F
\rangle_l (f)$, depending on the fixed (average) number of cavities
$l$ and the occupied fraction $f$. The possible values of $f$ are
extrapolated to the real interval $[0,l]$.  The expression for
$\langle \De F \rangle_l (f)$ is given by
\begin{eqnarray}
\langle \De F \rangle_l(f) &  = & \sum_{\NHT = 0}^N  S(\NHT;l,f) \label{DeF(f)} 
\end{eqnarray}
with 
\begin{equation}
\label{Gl:Sfaktor}
S(\NHT;l,f)    =  \sum_{X=0}^{\NHT} \sum_{Y=0}^{N-\NHT} 
\Om(\NHT,X,Y)\De F(\NHT) 
\end{equation}
and
\begin{equation}
\De F(\NHT)    =  - \frac{1}{\beta}\left[  B(\al,\om;\NHT)  X  +  B(\om,\al;\NHT)  Y  \right] .
\end{equation}
For the summation over the macroscopic parameters $\NHT$, $X$ and $Y$
the corresponding degeneracy (density)
\begin{equation}
\Om(\NHT,X,Y) = \frac{1}{4^N}\binom{N}{\NHT}
\binom{\NHT}{X} \binom{N-\NHT}{Y}  
\end{equation}
of microscopic configurations $\sT$ with respect to the hydrophobicity
$\NHT$ has to be taken into account.  Using the selectivity
(\ref{DeF(f)}) we consider the correction factor $ C_l(f) :=
\frac{\left\langle \De F \right\rangle_l (f)}{\left\langle \De F
  \right\rangle_l (f=0)}$ which relates the probability for having a
hydrated interface with $f\neq 0$ to the one for a dry interface with
$f=0$ (see section \ref{sec:enhanced}).

For sufficiently large $N$ a good approximation for $\left\langle \De
  F \right\rangle_l (f)$ can be obtained if we estimate the sums in
(\ref{DeF(f)}) by evaluating the strongly peaked function
$\Om(\NHT,X,Y)$ at its maximum
$\Om(\frac{N}{2},\frac{N}{4},\frac{N}{4})$. This fact may facilitate
future calculations, since for in the presented context ($N\approx
30\ldots 60$) there is almost no difference between the exact and the
approximated results. In figure \ref{fig:neuA-s04} the correction
factor $C_l(f)$ which is discussed in subsection \ref{sec:enhanced} is
shown for the exact average together with the approximation.

The selectivity (\ref{DeF(f)}) involves an average over all target
structures which are equally likely (expressed in terms of an average
over the corresponding hydrophobicities). For the investigation of the
optimal degree of hydration we are also interested in an analysis of
the free energy difference for a given fixed (averaged) hydrophobicity
$\langle \nHT \rangle$ of the target (see section \ref{sec:dry}). To
this end an additional Lagrange multiplier $\zeta$ that controls the
hydrophobicity of the target molecules has to be introduced. Note that
similarly the hydrophobicity of the rival has to be fixed by a
Lagrange parameter. As long as we choose the target and the rival to
have the same hydrophobicity, however, the results discussed below will not
depend on the hydrophobicity $\NHR$ of the rival. Hence, we replace
the probability $\propto \binom{N}{\NHT}$ for a configuration to have
the hydrophobicity $\NHT$ by the modified probability
\begin{equation}
P _\zeta(\NHT) = \frac{\exp ( - \zeta \NHT
  )}{(1+\exp(-\zeta))^N} \binom{N}{\NHT}.
\end{equation}
which can be used to express $\zeta$ in terms of a given $\langle \nHT
\rangle$.  Using this probability finally leads to a modified
correction factor $C_l(\langle \nHT \rangle;f)$ for given averaged
hydrophobicity $\langle \nHT \rangle$ of the target molecules:
\begin{equation}
C_l(\langle \nHT \rangle;f) = \frac{ \sum\limits_{\NHT=0}^N 
\left( \frac{\langle \nHT \rangle}{1- \langle \nHT \rangle} \right)^{\NHT}
S(\NHT;l,f)}{\sum\limits_{\NHT=0}^N 
\left( \frac{\langle \nHT \rangle}{1- \langle \nHT \rangle}\right)^{\NHT} S(\NHT;l,f=0)} ,
\label{Cmod}
\end{equation}
where $S(\NHT;l,f)$ is the function (\ref{Gl:Sfaktor}) of $\NHT$ and
$(l,f)$.

\acknowledgments

We gratefully thank Friederike Schmid for many fruitful discussions
and critically reading the manuscript.  This work was funded by the
Deutsche Forschungsgemeinschaft (SFB 613).

\nocite{*}
\bibliography{litpaper}

\end{document}